\documentclass[conference]{IEEEtran}
\IEEEoverridecommandlockouts
\usepackage{cite}
\usepackage{amsmath,amssymb,amsfonts}
\usepackage{algorithmic}
\usepackage{graphicx}
\usepackage{algorithm}
\usepackage{caption}
\usepackage{algorithmic}
\usepackage{textcomp}
\usepackage{xcolor, soul}
\usepackage{subfigure}
\def\BibTeX{{\rm B\kern-.05em{\sc i\kern-.025em b}\kern-.08em
    T\kern-.1667em\lower.7ex\hbox{E}\kern-.125emX}}
\usepackage{babel}
\begin{document}

\title{The Benefit of Decoder-Provided Pilots in Highly Dynamic Channels\\
\thanks{This material is based on work supported by the National Science Foundation (NSF) under grant ECCS-2433996 and the U.S. Air Force Research Laboratory (AFRL) under grant FA8650-23-2-1089. The U.S. Government is authorized to reproduce and distribute reprints for governmental purposes, notwithstanding any copyright notation thereon. The views and conclusions contained herein are those of the authors and should not be interpreted as necessarily representing the official policies or endorsements, either expressed or implied, of the United States Air Force, the Air Force Research Laboratory, or the U.S. Government.}
}

\author{\IEEEauthorblockN{
{Duschia Bodet}\IEEEauthorrefmark{1},
{Muriel M\'edard}\IEEEauthorrefmark{2},
{Muralidhar Rangaswamy}\IEEEauthorrefmark{3},
{Ken Duffy}\IEEEauthorrefmark{4}
}
\vspace{5pt}
\IEEEauthorblockA{\IEEEauthorrefmark{1}
Department of Electrical and Computer Engineering\\
Northeastern University, Boston, MA, USA 
}
\vspace{5pt}
\IEEEauthorblockA{\IEEEauthorrefmark{2}
Department of Electrical Engineering and Computer Science\\
MIT, Boston, MA, USA}
\vspace{5pt}
\IEEEauthorblockA{\IEEEauthorrefmark{3}
Sensors Directorate\\
Air Force Research Laboratory,WPAFB, USA}
\vspace{5pt}
\IEEEauthorblockA{\IEEEauthorrefmark{4}
Department of Mathematics \& Department of Electrical and Computer Engineering\\
Northeastern University, Boston, MA, USA}
}

\maketitle
\vspace{-2cm}
\begin{abstract}
Communications in highly dynamic channels relying on training-based channel estimation experience a trade-off between increasing channel measurement accuracy by sending more frequent training sequences and increasing data rate by sending fewer training sequences. Simultaneously, most communication systems use forward error correction to enable error detection and correction at the receiver. This paper presents decoder-provided pilots for time-varying channels by using decoded codewords as training sequences to update the channel estimate at the receiver. In contrast to approaches such as data-aided channel estimation, decision-feedback equalization, joint channel estimation and error correction, and turbo equalization, the decoder-provided pilots approach is non-iterative, which is ideal for low-latency requirements in highly dynamic scenarios. Furthermore, it is modulation-, code-, and decoder- agnostic, meaning it can be implemented on top of virtually any communication system that uses forward error correction. From an information-theoretic perspective, we derive the fundamental limits of decoder-provided pilots' ability to simultaneously sense the channel and transmit data. Simulation results demonstrate that decoder-provided pilots significantly improve performance, that when coding across frequency, soft-output can further enhance performance, and that when coding across time, short codes can outperform long codes of the same rate in fast-fading channels.
\end{abstract}


\section{Introduction}
\label{sec:introduction}
Estimating and tracking the channel is crucial for communications in highly dynamic channels. However, when relying on training-based channel estimation techniques, 
increasing the frequency of training improves the bit error rate (BER) but lowers the data rate. Furthermore, for communications in the millimeter wave and terahertz bands, channel coherence times will drop both due to an increased Doppler spread with the higher frequencies~\cite{rappaport2002wireless} as well as higher phase noise for terahertz superheterodyne receivers~\cite{sen2020teranova}. Both of these challenges can be addressed by more frequent training-based channel estimation, which sacrifices data rate.

A number of approaches have been taken to estimate and track the wireless channel while minimizing the required overhead for training. Broadly, the approaches fall into two categories: those attacking it from a modulation/demodulation perspective and those attempting to optimize the arrangement of pilot or training symbols to send the fewest possible. We describe these approaches in more detail in the next section, but they all ignore one significant advantage.

Almost every wireless communication system uses forward error correction (FEC) to add redundant bits that allow the receiver to detect and correct errors. Several works have demonstrated ways to exploit this structure to perform joint coding and channel estimation~\cite{al2003exploiting,lopes2001exploiting,wiame2025joint,ghaddar2021joint,su2002low, douillard1995iterative,otnes2005iterative,laot2002turbo,tuchler2002turbo}. Although (unlike the modulation-based approaches) joint coding and channel estimation techniques allow for flexibility in the choice of modulation, they are code- or decoder-specific. Thus, they can only be implemented when a specific code or decoder is used, which limits their flexibility. 

Furthermore, most of these schemes are iterative, which means they achieve performance gains by sacrificing latency. Thus, they may not be ideal for applications requiring ultra-reliable low-latency communications (URLLC). Some data-aided channel estimation techniques also take advantage of the decoding, but these are often also iterative~\cite{ma2014data,yuan2014soft} or rely on deep learning~\cite{jeon2020data,hashempoor2025deep}.

In this paper, we propose using decoded codewords as training sequences to reduce the required number of designated training sequences transmitted and thereby increase the throughput. Not only is our approach modulation-agnostic and therefore \textit{can be used to accurately track the channel using any modulation scheme}, but it also only requires the hard-decision output of a decoder, which makes it \textit{code- and decoder-agnostic} as well. Moreover, we show that it outperforms comparable modulation-level approaches like the decision-feedback equalizer (DFE) for higher order modulations. Furthermore, the approach is \textit{non-iterative} and useful in highly volatile environments requiring low latency. 

We offer variations of the proposed algorithm to account for single- and multi-carrier systems and to take advantage of cyclic redundancy checks (CRCs) or soft-output. We show that using the soft output of does not change the performance for single-carrier waveforms, but can substantially improve the performance for multi-carrier systems. 

Specifically, the contributions of this paper are as follows: 
\begin{enumerate}
    \item \textbf{We present decoder-provided and CRC-gated decoder-provided pilots for channel estimation in fast-fading channels using hard decision output only.} The proposed algorithm utilizes decoded codewords as additional pilot sequences to enhance channel estimation. Decoder-provided pilots are pilots generated from the decoded codeword regardless of whether the receiver is confident in the decoding, while CRC-gated decoder-provided pilots are generated only when a CRC checksum returns zero.
    \item \textbf{We analytically demonstrate the limits of a communication systems in dynamic channels as the interval between channel estimates increases and how these limits are impacted by the length of the codeword.} Although it has been shown that for a given code rate increasing the codeword length allows the performance to approach capacity~\cite{shannon1948mathematical}, we show that in fast fading channels this approached capacity is  dramatically reduced if the codeword length approaches the channel's coherence time. 
    \item \textbf{We validate the performance of the proposed algorithm and the implications of the derived results through extensive simulations.} The simulation results demonstrate significant performance gains by utilizing decoder-provided pilots in various fast fading channels for single-carrier and multi-carrier waveforms. 
    \item \textbf{We demonstrate that utilizing soft-output enhances performance when coding across frequency but not time, and that decoder-provided pilots are preferred for systematic codes while CRC-gated decoder-provided pilots are better for non-systematic codes.} This contribution both offers important notes for practical implementation and emphasizes the advantage of decoder-provided pilots offering full enhancements without any soft-output information when coding across time. 
\end{enumerate}

In the following section we discuss related works. Then, in Sec.~\ref{sec:system-model}, we describe the proposed algorithms. Our analytical analysis is in Sec.~\ref{sec:analysis} and simulated results are in Sec.~\ref{sec:results}. Final discussion and conclusions are provided in Sec.~\ref{sec:conclusion}.

\section{Related Work}
\label{sec:related-work}
\subsection{Differential Demodulation and Detection} 
One common approach for eliminating or reducing the requirement of accurate channel state information at the receiver is to use differential demodulation, where information is embedded in the difference in phase between the prior and current symbol. This approach is used in the IEEE IS-54 standard, and has been extended to be utilized for multiple-input multiple-output (MIMO) systems through space-time codes (STCs)~\cite{tarokh1998new,tarokh2002differential}. Unlike error correcting codes which are implemented (usually at the bit level) to identify and potentially correct bit errors in transmission, STCs are implemented at the symbol level to take advantage of spatial diversity and maximize the signal-to-noise ratio (SNR). In~\cite{tarokh1998new}, the authors use STCs to reduce and even eliminate the need for pilot training sequences by performing differential detection. More recently in~\cite{mattu2025differential}, the authors propose a differential communication scheme using Zak-OTFS, one of the candidate waveforms for 6G.

Although differential detection has the potential to eliminate the need for training or pilot transmissions it limits the system's choice of modulation (or multi-antenna processing in the case of diversity) to options that enable differential detection. Furthermore, pilot symbol-assisted modulation has been shown to outperform differential detection~\cite{cavers2002analysis}.

\subsection{Data-Aided Channel Estimation \& Decision Feedback Equalization}
Using data sequences as additional pilot sequences in channel estimation has been implemented in data-aided channel estimation, where symbols or subcarriers that have been detected with high confidence are used as pilots. This approach can either learn which symbols are reliable, sometimes through sparse Bayesian learning~\cite{prasad2014joint} or deep learning~\cite{jeon2020data,hashempoor2025deep} or use soft-output from a decoder~\cite{yuan2014soft}. However, these approaches, including the approach in~\cite{yuan2014soft} is iterative, which means it may not be ideal for low-latency applications or highly dynamic scenarios. 

Decision feedback equalization (DFE) uses knowledge of the channel to reduce the impact of inter-symbol interference of a multi-tap channel by subtracting previously demodulated symbols from the current symbol according to the channel impulse response~\cite{belfiore1979decision,cioffi1995mmse,george1971adaptive}. DFEs, however, can suffer from severe error propagation.
\subsection{Engineering Pilot Arrangement at the Transmitter}
One crucial consideration is placing training symbols among data transmissions to optimize performance. In~\cite{coskun2005combined}, the authors approach this challenge as a coding question with a code rate of $(N-K)/N$, where $N$ is the number of data bits, $K$ is the number of pilot bits, and $K < N$. They present an algorithm to determine the minimum amount of overhead required as well as an algorithm to design corresponding ``training codes" to enhance the performance. However, their approach struggles to scale for long data frames. The authors of~\cite{dong2002optimal} determine more broadly the ideal placement of pilot symbols in a frame for channel estimation. For OFDM systems, the authors of~\cite{coleri2002channel} compare different arrangements of pilot symbols within an OFDM block and interpolate methods of the channel calculated by those pilots.

\subsection{Exploiting FEC for Channel Estimation}
Nearly all of these approaches ignore the advantages available through FEC. Since communication systems implement some form of FEC, the aforementioned approaches would most probably be implemented with a FEC code on top of them, which could provide additional gains. A number of works have endeavored to utilize the structure provided by FEC to improve channel estimation. 

\subsubsection{Turbo Equalization}
First presented in~\cite{douillard1995iterative} and expanded upon substantially since~\cite{tuchler2002turbo,koetter2004turbo,laot2002turbo}, turbo equalization accounts for inter-symbol interference (ISI) by using soft-output detection and decoding to iteratively improve the decoding. This concept has been extended to use soft-output information to improve the decoding and refine a channel estimate~\cite{valenti2002iterative, otnes2005iterative,al2003exploiting}. Turbo equalization achieves impressive performance, which has led to a number of adjacent algorithms proposed~\cite{lopes2001exploiting,qin2023joint}, including approaches that combine turbo-equalization with differential detection~\cite{hoeher2002turbo,marsland2002multiple}. However, the turbo equalization process is generally embedded into the decoding process, so the ability to utilize the decoding in the channel estimation depends on the code and decoder used. 

\subsubsection{Other Joint Channel Estimation and Decoding Approaches}
A few strategies for joint channel estimation and (de)coding have been explored for polar codes~\cite{ghaddar2021joint}. The authors use a list decoder to implement an algorithm where bits that are decoded with high reliability are used as additional pilot symbols. Since this work uses a specific decoder for polar codes - the successive cancellation trellis decoder - in the system design, the algorithm is specific to polar codes. 

The authors of~\cite{su2002low} perform joint channel estimation and decoding in an iterative process using a soft-input soft-output decoder, and in~\cite{al2003exploiting} a similar approach is shown for OFDM systems. However both these approaches rely on soft-output information from the decoder and multiple iterations to achieve improved performance.

In~\cite{wiame2025joint}, the authors use ORBGRAND~\cite{duffy2022ordered}, a Guessing Random Additive Noise Decoder (GRAND)~\cite{duffy2019capacity} to take the channel estimation error into consideration. All decoders in the GRAND family are universal decoders. Therefore, the algorithm presented in~\cite{wiame2025joint} is code-agnostic, but it is decoder-specific as it relies on ORBGRAND. \\

All these approaches offer flexibility at the modulation level, but enforce some restrictions at the coding level that limit their versatility. 

In addition to being modulation-agnostic, ur presented algorithm is code- and decoder-agnostic, which allows for easy adaptation in many communication systems. Additionally, decoder-provided pilots provide a non-iterative alternative to joint decoding and channel estimation techniques which can reduce latencies. 

\section{System Model}
\label{sec:system-model}
\subsection{Decoder-Provided Pilots}
\label{sec:system-model-long-codes}
\begin{figure*}[!h]
    \centering
    \includegraphics[width=0.8\linewidth]{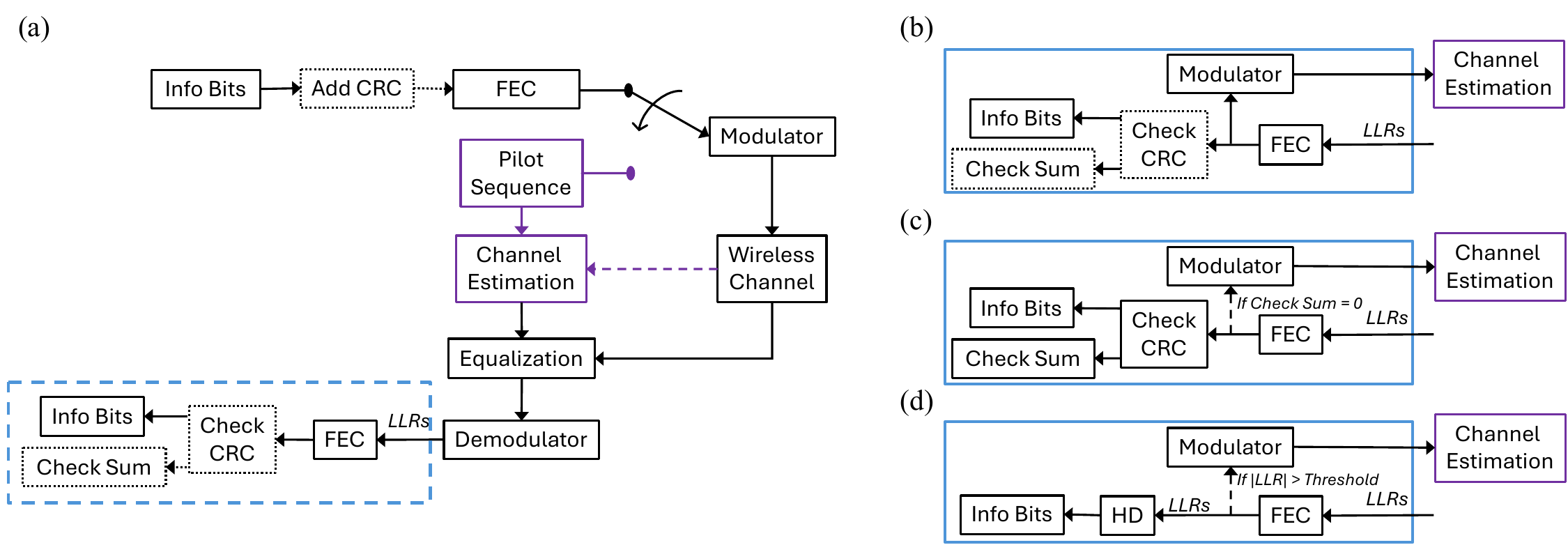}
    \caption{System model showing the block diagram for the system: (a) only relying on designated training sequences, (b) using decoder-provided pilots, (c) using CRC-provided pilots, and (d) using reliability-thresholded decoder-provided pilots.}
    \label{fig:system-model-long-codes}
\end{figure*}
A traditional wireless communication system is shown in Fig.~\ref{fig:system-model-long-codes}a. We assume a binary data sequence $U$ of length $k$ is optionally supplemented with a CRC to generate the sequence $U_{CRC}$ of length $n_{CRC}$ (Unless otherwise specified in this paper, capitalized variables denote vectors). The CRC can be added to enable error detection after forward error correction at the receiver. Thus $U_{CRC}$ (or $U$ if no CRC is added) is passed to the FEC block, which maps the sequence to a codeword $C$ of length $n$. The codeword is modulated, resulting in the vector $X$, before being transmitted over the wireless channel.

Equalization follows using channel estimates obtained through prior pilot training, and then demodulation and decoding are performed to obtain the information stream. After decoding, the CRC checksum is computed if a CRC is used, otherwise the decoding result is directly returned as information bits. For our analysis, we assume that known codewords are sent at some training interval, $t_I$. Such that every $t_I / (\frac{n}{\log_2 (M)} T_s)$ transmitted block is the designated pilot training sequence, where $M$ is the modulation order and $T_s$ is the symbol duration.

We present two approaches to enable data blocks to be used as pilot blocks: (1) decoder-provided pilots and (2) CRC-gated decoder-provided pilots. 

\subsubsection{Decoder-Provided Pilots}
Replacing the dashed box in Fig.~\ref{fig:system-model-long-codes}a with Fig.~\ref{fig:system-model-long-codes}b, we achieve a system using what we decoder-provided pilots. In this case, the decoded sequence is always used as an additional pilot sequence regardless of any CRC checksum's outcome. We consider this case because even for the purposes of channel estimation, one or a few bit errors out of the entire transmitted codeword will likely lead to minimal degradation in the channel estimate. In other words, we expect the error in channel estimation due to a few bit errors to be much less than the error in channel estimation due to a stale channel measurement. 

\subsubsection{CRC-Gated Decoder-Provided Pilots (i.e., ``CRC-provided pilots"} 
Similarly, the second approach is depicted in Fig.~\ref{fig:system-model-long-codes}c. In this case, the CRC must return zero in order for the decoded sequence to be utilized as an additional pilot sequence. Essentially, we use the CRC to flag decoded data blocks that the receiver is confident are correct and use these blocks for additional pilot training. 

\subsubsection{Reliability-Thresholded Decoder-Provided Pilots for Coding Across Frequency}
When coding across frequency, we can utilize the soft-output of individual IQ symbols on each subcarrier to determine whether a decoded symbol is used for channel estimation. In this case, we make the same assumptions as described above except that the length of the codeword in symbols is equal to the number of data subcarriers (i.e., $\frac{n}{\log_2(M)} = N_{sbc}$; each codeword corresponds to one OFDM symbol) and training blocks are also OFDM symbols. Thus every $t_I / T_s$ transmitted block is a pilot training sequence with $T_s$ being the duration of an OFDM symbol. We assume a time-varying channel with coherence bandwidths on the order of the subcarrier spacing, which necessitates tracking the channel on each subcarrier individually. Each OFDM symbol corresponds to a single codeword, and therefore, each IQ symbol within each OFDM symbol corresponds to a single subcarrier. If soft-output information is available for each IQ symbol, we can use the soft-output information to decide whether we use the decoded IQ symbol to update our channel estimate or keep our original channel estimate. In other words, we only update our channel estimate on subcarriers for which the reliability exceeds a certain threshold, as illustrated in Fig.~\ref{fig:system-model-long-codes}d.

When coding across time, the symbol or bit-wise soft-output is less helpful since channel estimation is performed block-wise. To demonstrate this, we present a two approaches for utilizing soft-output from the list decoder for comparison purposes:

\textit{1) Weighted IQ Average Decoder-Provided Pilots:} In this approach, the pilot sequence provided by the decoder is generated by taking a weighted average of the IQ symbols of each potential codeword according to their probabilities. More specifically, each of the $L$ potential codewords, $\hat{C_l}$, are modulated to generate an IQ stream, $\hat{X_l}$. The pilot sequence is then given by $P = \sum_{l=1}^L \frac{p_l \: \hat{X_l}}{\Sigma_{l=1}^L p_l}$,
where $p_l$ is the probability that the $l$th codeword is the correct codeword. 

\textit{2) Minimizing Change in Channel Estimate Decoder-Provided Pilots:} This approach focuses on minimizing the change in the channel estimate as a method of decision-making. Essentially, each potential codeword, $\hat{C_l}$, is modulated to determine $n$ IQ streams, $\hat{X_l}$. The IQ streams are all fed to the channel estimation block that has also been provided with the received signal $Y$. A list of candidate channels for the current codeword $\hat{F}_l (i)$ are determined, and the pilot sequence used for the subsequent codeword is given by: $P = \arg \min_{l} \bigg| \hat{F_l}(i) - \hat{F_l}(i-1) \bigg|$
In other words, the potential codewords are all used to determine candidate channels, and the codeword that minimizes the change in the channel estimate based on the previous channel measurement is chosen.

\subsection{Effective Information Rate} 
As a metric of the system performance, we calculate the effective information rate of the system, which we define as the number of correctly received information bits divided by the total number of bits transmitted. It is given by: 
\begin{equation}
    R_{eff} =  (1 - BLER)\frac{T_I - 1}{T_I}\frac{k}{n}, 
    \label{eq:eff-rate}
\end{equation}
where $T_I = t_I / (\frac{n}{\log_2 (M)} T_s)$. Thus, $T_I - 1$ is the number of data blocks transmitted in one interval, $T_I$ is the total number of blocks (including pilot blocks), and $BLER$ is the block error rate.

\section{Analytical Performance}
\label{sec:analysis}
In this section, we analytically derive the performance of the systems utilizing decoder- and CRC-provided pilots. 

We assume the system described in Sec.~\ref{sec:system-model} experiences a wireless channel such that the received sequence is given by 
\begin{equation}
    Y = FX + N,
\end{equation}
where $F$ is the measured channel, $N$ represents additive white Gaussain noise, and $X$ is the transmitted sequence. In the case of imperfect channel estimation, the received signal can be represented as
\begin{equation}
    Y = \bar{F}X + \tilde{F}X + N,
    \label{eq:y=fx+fx+n}
\end{equation}
where we have defined $F = \bar{F} + \tilde{F}$, with $\bar{F}$ being the estimated channel response and $\tilde{F}$ being the error in the channel estimate. 

As first shown in~\cite{shannon1948mathematical}, the rate of communication will be bounded by the mutual information, 
\begin{equation}
    R < H(X) - H(X|Y).
    \label{eq:r-bound}
\end{equation}
In~\cite{medard2002effect} it was shown that for the system described by Eq.~\ref{eq:y=fx+fx+n}, the differential entropy $H(X|Y)$ will be upper bounded by 
\begin{equation}
    H(X|Y) \leq k\ln(2\pi e) - \frac{1}{2}\ln\big( | \Lambda_{X}^{-1} + \bar{F}^{T} (\Lambda_{\tilde{F}X} + \Lambda_{N})^{-1} \bar{F}| \big),
    \label{eq:h-bound}
\end{equation}
where $k$ is the length of the sequence and $\Lambda_A$ designates the covariance of a random variable $A$. We assume $\Lambda_X$ and $\Lambda_N$ (i.e., the covariance of the source and noise, respectively) are known. We also assume knowledge of the channel characteristics. Thus, we are left to find $\Lambda_{\tilde{F}X}$. Since the source and the channel estimation error are independent, the off-diagonal elements of $\Lambda_{\tilde{F}X}$ will be $0$. We have already established knowledge of the source's covariance, so the only remaining mystery is the variance of $\tilde{F}$, $\sigma^2_{\tilde{F}}$. 

For fast fading channels, $\sigma^2_{\tilde{F}}$ can be in one of three states: 
\textbf{State 1.} In the case where we have a fairly reliable channel estimate, $\tilde{F}$ can be represented as a zero-mean Gaussian random variable with a low variance,
\begin{equation}
    \sigma^2_{\tilde{F}_{state1}}=\sigma_e^2.
\end{equation}
This scenario corresponds to $t  << \tau_{C}$, where $\tau_{C}$ is the coherence time of the channel and $t$ is the time since the most recent channel estimate. (This state is identical to the case studied in~\cite{medard2002effect}.)\\
\textbf{State 2.} As $t$ approaches $\tau_{C}$, the measured channel $F$ is somewhat correlated to the true channel $\bar{F}$, but not so closely that $\tilde{F}$ can be represented as a zero-mean Gaussian. In this case, $\tilde{F} = F - \bar{F}$, thus $\sigma^2_{\tilde{F}} = \sigma^2_{F} + \sigma^2_{\bar{F}} - 2\Lambda_{F, \bar{F}} $. Since $F$ and $\bar{F}$ are two instances of the same random variable
\begin{equation}
    \sigma^2_{\tilde{F}_{state2}} = 2\sigma^2_{F} -2\Lambda_{F, \bar{F}}
    \label{eq:var-f-tilde-stage-2}
\end{equation}
\textbf{State 3.} For $t >> \tau_{C}$, $\bar{F}$ and $F$ are no longer correlated, sending $2\Lambda_{F, \bar{F}} \longrightarrow 0$, and leaving
\begin{equation}
    \sigma^2_{\tilde{F}_{state3}} = 2\sigma^2_{F}
\end{equation}   

It is important to recall that the coherence time, $\tau_C$, indicates the duration over which the channel is expected to be highly correlated. Durations greater than $\tau_C$ will still share some correlation that is less than $0.5$. Therefore, as $t$ increases from $0$ to $\infty$, $\tilde{F}$ will move gradually from State 1, through State 2, and into State 3. Additionally, $\sigma^2_{\tilde{F}_{state1}} < \sigma^2_{\tilde{F}_{state2}} < \sigma^2_{\tilde{F}_{state3}}$, so the variance of the channel estimation error will monotonically increase with time. 
When used in Eq.~\ref{eq:h-bound}, a larger $\sigma^2_{\tilde{F}}$ raises the upper bound on the differential entropy, leading to less mutual information in Eq.~\ref{eq:r-bound} and a lower system capacity. 

In practice, channel estimation is performed periodically to prevent $\sigma^2_{\tilde{F}}$ from reaching stages 2 and 3. Thus, the data rate of the system in bits per channel use is given by 
\begin{equation}
    R = R_{code} \bigg(1 - \frac{1}{T_I}\bigg)\log_2(M),
    \label{eq:r-eff}
\end{equation}
where $R_{code}$ is the code rate and $M$ is the modulation order. The data rate in Eq.~\ref{eq:r-eff} is bounded by Eq.~\ref{eq:r-bound}.
\begin{equation}
    R_{code} \bigg(1 - \frac{1}{T_I}\bigg)\log_2(M) < H(X) - H(X|Y),
    \label{eq:r-eff-bounded}
\end{equation}
where $T_I$ is generally chosen such that $t_I << \tau_{C}$. There is an apparent trade-off between increasing the $t_I$ to enhance the channel estimate (which lowers $H(X|Y)$ and thereby increases the capacity) and decreasing $t_I$ to increase the data rate. If $t_I$ approaches or exceed $\tau_{C}$, $H(X|Y)$ would increase with the $\sigma^2_{\tilde{F}}$ to limit the capacity, as described above. However, increasing $T_I$, increases the left-hand side of Eq.~\eqref{eq:r-eff-bounded} to allow the data rate to approach the capacity. 

This trade-off can also be considered as a choice between operating in a channel that is the average of many various states or operating in a channel with greater knowledge of its current state. Unless $\sigma_{F}^2$ is quite low, Eq.~\eqref{eq:h-bound} demonstrates that the former will generally allow for higher achievable capacities.\\

The objective of the proposed decoder-provided pilot systems is to allow $t_I$ to exceed the coherence time while keeping the $\sigma^2_{\tilde{F}}$ to a minimum (i.e., without lowering the channel capacity). 

\subsection{CRC-Provided Pilots}
\label{sec:analysis-crc-provided}
We assume that the CRC is designed such that it provides an accurate indication of whether a codeword is detected in error. Thus, following the procedure described in Sec.~\ref{sec:system-model}, the CRC-provided pilot sequences will be used only if the codeword is correctly decoded. 

The probability of correct decoding is given by~\cite{shannon1948mathematical}
\begin{equation}
    P_c = (1 - 2^{n(R_{code}-H(X))})^{2^n H(X|Y)},
    \label{eq:prob-correct-cw}
\end{equation}
where from Eq.~\eqref{eq:r-eff-bounded}
\begin{equation}
    R_{code} < \frac{H(X) - H(X|Y)}{\big(1 - \frac{1}{t_I}\big)\log_2(M)}.
    \label{eq:r-code-bound}
\end{equation}
$H(X|Y)$ is still bounded by Eq.~\ref{eq:h-bound} and will follow $\sigma^2_{\tilde{F}}$. 

When utilizing CRC-provided pilots to supplement designated training sequences sent every $t_I \sim \tau_{c_{blocks}}$ blocks, the variance of $\tilde{F}$ is given by Eq.~\ref{eq:var-f-tilde-stage-2}. 

If we assume the channel to be a Gauss-Markov process and that the measured channel $\bar{F}$ is an accurate representation of a previous channel state (i.e., $\bar{F}(i_m) =F(i_m)$), then
\begin{equation}
    \bar{F}(i, i_m) = \beta^{i-i_m+1} F(i_m) + \sum_{j=0}^{i-i_m} \beta^j Z(i - i_m -j), 
\end{equation}
where $i_m$ is the time instance when $F(i_m)$ was measured, $\beta$ is a constant related to the coherence time as in~\cite{medard2002effect}, and each $Z$ is an independent Gaussian random variable with variance $\sigma_Z^2 = (1 - \beta^2)\sigma_F^2$. (We consider $\beta$ to be a constant for simplicity, but $\beta$ can also be a function of $i$ to change with time and/or $\beta$ can change with frequency if the signaling is occurring across frequency.) 

The covariance will be given by, 
\begin{equation}
\begin{split}
    \Lambda&_{F(i, i_m),\bar{F}(i_m)} = 
    \\ & E\bigg[F(i_m) \bigg( \beta^{i-i_m+1} F(i_m) + \sum_{j=0}^{i-i_m} \beta^j Z(i - i_m -j) \bigg)\bigg] \\ &- E[F]^2,
\end{split}
    \label{eq:cov-f-fbar}
\end{equation}
where $E[.]$ is the expectation of a random variable, and we have used the fact that $F$ and $\bar{F}$ are both instances of a random variable with the same distribution, thus $E[F] = E[\bar{F}]$.

When solely relying on designated training sequences every $T_I$ blocks, the covariance becomes
\begin{equation}
\begin{split}
    \Lambda&_{F(i, i_m),\bar{F}(i_m)} = 
    \\ & E\bigg[F(i_m) \bigg( \beta^{i-i_m+1} F(i_m) + \sum_{j=0}^{\text{mod}(i-i_m,T_I)} \beta^j Z(i - i_m -j) \bigg)\bigg] \\ &- E[F]^2,
\end{split}
\label{eq:cov-t_i}
\end{equation}

With the CRC-provided pilots, the final index of the summation in Eq.~\eqref{eq:cov-t_i} covariance becomes 
\begin{equation}
\begin{split}
    \text{mod}\bigg(i-i_m,\text{min}(T_I, \frac{1}{P_c(i-1)})\bigg),
\end{split}
\end{equation}

In the unlikely event that $P_c \longrightarrow 0$, there will be no additional pilot sequences, and $\sigma^2_{F}$ will follow Eq.~\ref{eq:var-f-tilde-stage-2}. This means that CRC-provided pilots will never hurt the performance of the system; they only have the potential to improve it. 

As $P_c \longrightarrow 1$, the channel estimate will be updated every block, leading to $i_m = i - 1$ and the covariance given by
\begin{equation}
\begin{split}
    \Lambda&_{F(i, i-1),\bar{F}(i-1)} = 
    \\ & E\bigg[F(i-1) \bigg( \beta^{2} F(i-1) + \beta Z(0) + Z(1) \bigg)\bigg] \\ &- E[F]^2,
\end{split}
\end{equation}
which can be rearranged, utilizing the linearity of the expectation:
\begin{equation}
\begin{split}
    \Lambda_{F(i, i-1),\bar{F}(i-1)} = \\&E\big[\beta^2\bar{F}^2(i-1)\big] + E\big[\beta\bar{F}(i-1)Z(0)\big]\\& + E\big[\bar{F}(i-1)Z(1)\big] - E[F]^2.
\end{split}
\end{equation}
$Z$ and $\bar{F}$ are independent, and $Z$ is zero-mean, which eliminates the second term. Leaving: 
\begin{equation}
\Lambda_{F(i, i-1),\bar{F}(i-1)} = E\big[\beta^2\bar{F}^2(i-1)\big] - E[F]^2.
\end{equation}
which - except for the $\beta^2$ in the first term - resembles the definition of the variance. Thus, Eq.~\eqref{eq:var-f-tilde-stage-2} will be a small constant as in Stage 1, regardless of $t_I$. In other words, for $P_c$ close to 1, the dependence on $t_I$ is eliminated. Furthermore, given that many FEC schemes are required to operate with a probability of bit error on the order of $10^{-5}$ or $10^{-6}$, in many practical scenarios $P_c$ will be very close to 1. 

\subsection{Decoder-Provided Pilots}
\label{sec:analysis-decoder-provided}
For the decoder-provided pilots, there is no assumption that sequences are correctly decoded when they are used for channel estimation. Thus, for scenarios with low signal power or low $P_c$, the channel estimate will be incorrect because the decoded sequence $\hat{X} \neq X$. 

The probability of bit errors in the sequence $\hat{X}$ is bounded according to~\cite{moon2020error}
\begin{equation}
    \frac{1 - P_c}{n} \leq P_b \leq 1 - P_c.
    \label{eq:bit-error-bounds}
\end{equation}
Because we are using the entire sequence (i.e., not only the $k$ information bits of a codeword), we use $1/n$ instead of $1/k$ as in~\cite{moon2020error}. Thus, the chance of bit error can be smaller than if we were only concerned with information bits. In general, Eq.~\eqref{eq:bit-error-bounds} shows that increasing $n$ in the denominator leads to a looser lower bound (i.e., the potential for lower bit error rates). Furthermore, for a given code rate $R_{code}$, it has been shown~\cite{shannon1948mathematical} that as $n\longrightarrow\infty$, $1-P_c \longrightarrow0$. 

\subsection{Coherence Time's Limit on Codeword Length}
\label{sec:analysis-limit-length}
To avoid error propagation, the system clearly should operate in the regime where $1-P_c \longrightarrow 0$. According to~\cite{shannon1948mathematical} and Eq.~\eqref{eq:bit-error-bounds}, it may seem that making $n$ arbitrarily large for a given rate $R_{code}$ will lead to an increase in $P_c$, a decrease in $1/n$, and therefore, an improvement in the overall performance of the decoder-provided system. However, in the time-varying channel, letting $n\longrightarrow\infty$ will eventually lead to the case where $n$ approaches and exceeds the coherence time, where the channel becomes fast fading rather than block fading. In our prior analysis, we assume that the channel remains constant over the duration of a codeword. For time-varying channels, $n$ cannot approach $\infty$ with that remaining true.

As $n \longrightarrow \infty$ in the fast fading channel, the channel experienced by each codeword is no longer correlated, thus the variance will perpetually be in stages 2 and 3. $R_{code}$ will again be limited by Eq.~\eqref{eq:r-code-bound}, and $H(X|Y)$ will increase with increasing $\sigma_{\tilde{F}}$ as before, leading to a decrease in achievable capacity. 

\begin{figure}[h!]
         \centering
         \includegraphics[width=0.8\linewidth]{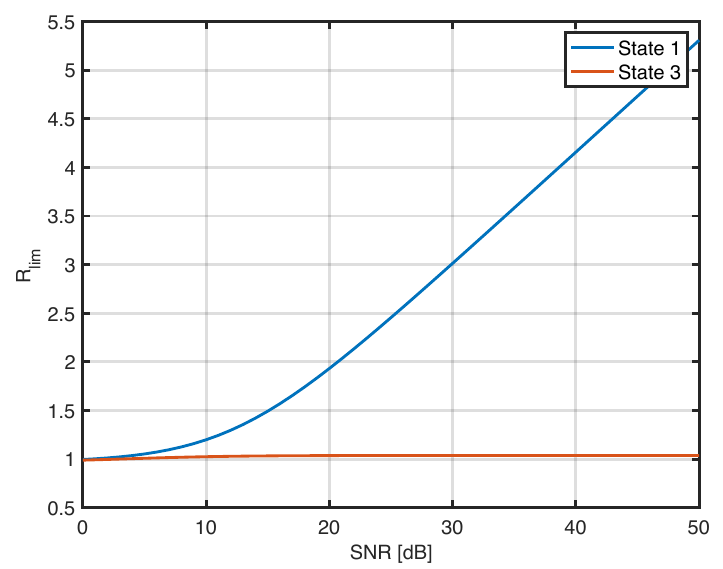}
    \caption{Capacity of time-varying channel using the upperbound for $H(X|Y)$ given by Eq.~\eqref{eq:h-bound}} 
    \label{fig:analytical-rate-limit}
\end{figure}

In Fig.~\ref{fig:analytical-rate-limit}, we plot the capacity bounds when the channel is in States 1 and 3 given by Eq.~\eqref{eq:r-bound} and~\eqref{eq:h-bound}, assuming a Rayleigh fading channel and a binary symmetric source. The capacity of State 2, would be between States 1 and 3, so this plot demonstrates the upper and lower bounds of the capacity. Despite increasing the signal-to-noise ratio (SNR), the capacity of the system with a stale channel estimate (i.e., a system in State 3) is severely limited compared to a system with up-to-date channel information (i.e., a system in State 1). Decoder-provided pilots allow us to remain in State 1, while reducing the number of transmitted pilot sequences. However, if the codeword length is on the order of the channel coherence time, this will push the system towards State 3. 

Therefore to maintain a higher capacity, the length of the codeword should be limited according to the coherence time if other methods such as Kalman filtering or inserting training symbols within codewords are not implemented. 

\section{Simulated Results}
\label{sec:results}
\subsection{Simulation Framework}
\label{sec:results-framework}
\noindent\textit{Channel Model}\\
\label{sec:chnl-model}
We assume an auto-regressive Rayleigh fading channel similar to the ones given by~\cite{abou2005binary,medard2002effect,sarieddeen2022grand}:
\begin{equation}
    \bar{F}(i) = \beta \bar{F}(i-1) + Z(i),
    \label{eq:rayleigh}
\end{equation}
as described in Sec.~\ref{sec:analysis}. The variance of $\bar{F}$, $\sigma_{\bar{F}}$, is set to 1, and the scaling factor $\beta$ can be calculated according to $\beta = \phi^{1/\tau_{C}W}$. Where the coherence time, the transmission bandwidth, and the level of correlation for which the coherence time is defined are represented as $\tau_C$, $W$, and $\phi$, respectively. In our analysis, we use a block fading channel, where the time index $i$ of each channel update is equal to the duration of a codeword.  

In the case of a multi-tap channel we specify a delay profile $D[i]$, which for a dicode channel takes the form: 
\[
 D[i] = \begin{cases}
     1 & i = 0\\
     \rho & i = 1\\
     0 & \text{otherwise}
 \end{cases}
\]  
where $\rho < 1$. 

We use the channel estimate ${F}$ to implement an MMSE equalizer: 
\begin{equation}
    Y_{MMSE}(i) = (F^* F + \sigma_n^2)^{-1} F^* Y(i)
\end{equation}
Thus, according to~\cite{sarieddeen2022grand}, the log-likelihood ratio (LLR) $L(i,j)$ for the $j$th bit of the $i$th symbol is calculated
\begin{equation}
\scriptstyle
    L(i,j) = \frac{\sum e^{-\left((Re\{Y(i)\} - Re\{x \in \chi^{(0)}\})^2 + (Im\{Y(i)\} - Im\{x \in \chi^{(0)}\})^2\right)/\sigma^2}}{\sum e^{-\left((Re\{Y(i)\} - Re\{x \in \chi^{(1)}\})^2 + (Im\{Y(i)\} - Im\{x \in \chi^{(1)}\})^2\right)/\sigma^2}},
    \label{eq:llrs}
\end{equation}
where $\chi^{(0)}$ and $\chi^{(1)}$ correspond to the symbols in the transmitted alphabet where the $j$th bit of the $i$th symbol are 0 and 1, respectively. The variance $\sigma^2$ in~\eqref{eq:llrs} is given by $\sigma^2 = \sigma_n^2 (F^*F + \sigma_n^2)^{-1}$ to account for the channel estimate. \\

\noindent\textit{Demodulator-Provided and Free Pilots}\\
\label{sec:demod-genie}
In addition to the presented algorithms, we also include demodulator-provided and free pilots in our analysis for comparison purposes. The demodulator-provided pilots are obtained by performing hard-decision detection on the output of the demodulator and passing this result to the channel estimation block without performing any decoding. This approach is essentially the decision-feedback equalizer (DFE)~\cite{belfiore1979decision} and is used as a benchmark. 

The free pilots are obtained by gifting the receiver a perfectly decoded previous transmission block. In other words, any data block in the free pilot system is received as if the previous block is a designated training block. Thus, the system using free pilots shows the best possible performance of the system.

\subsection{Decoder-Provided Pilots using Long Codes \& CRCs}
\label{sec:results-long-codes}
\begin{figure}[h!]
        \centering
        \includegraphics[width=\linewidth]{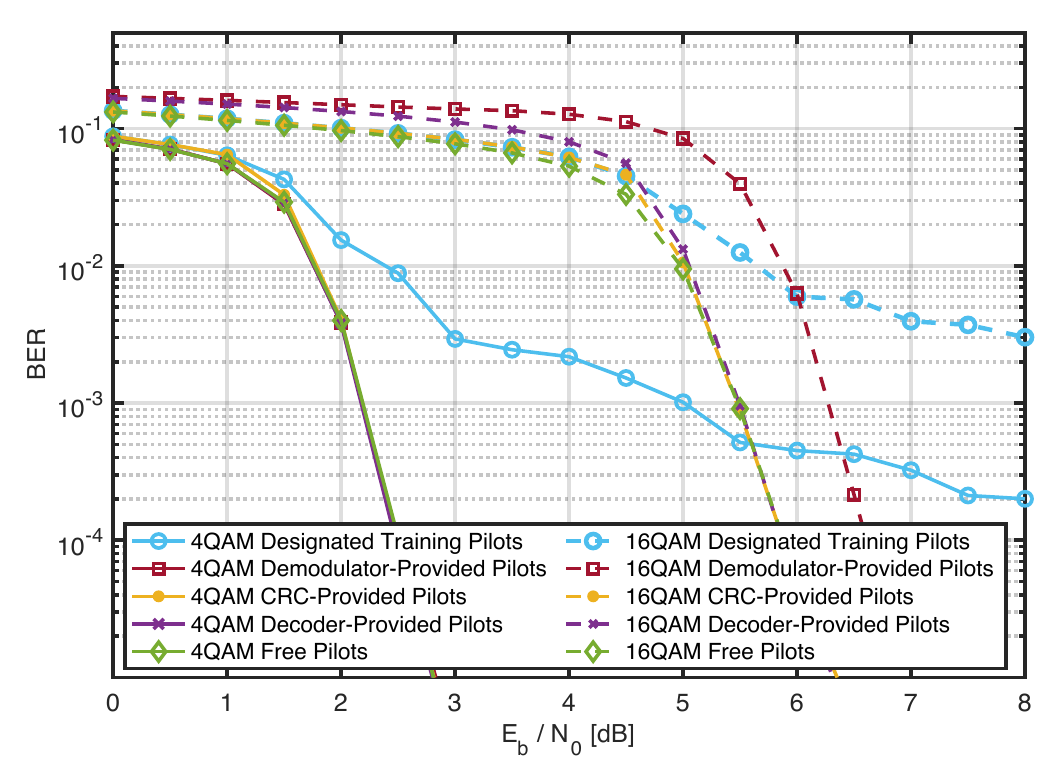}
        \caption{Bit error rate (BER) performance of a $eBCH^2 [1024, 676]$ component code with 11-bit CRC in a single-tap Rayleigh fading channel with $t_I = 100$ codewords and $D_I = 25$K symbols}
        \label{fig:mod-order-comparison-product-codes} 
\end{figure}

\noindent\textit{Decoder-Provided Pilots Reducing Error Propagation for High Modulation Orders}\\
The BER performance results of our initial analysis are shown in Fig.~\ref{fig:mod-order-comparison-product-codes} and utilize the hard decision output of a state-of-the-art GRAND decoder presented in~\cite{yuan2025soft}. We implement $eBCH [32,26]^2 = [1024, 676]$ component codes with an 11-bit CRC according to Koopman's polynomials~\cite{koopman2004cyclic}. A designated training sequence is transmitted every $t_I = 100$ codewords, and the decorrelation interval of the channel is given by $D_I = 25*10^{3}$, which means the channel decorrelates about every $25,000$ symbols, which is about equivalent to every $25$ codewords. In other words, $\tau_{C} / (\frac{nT_s}{\log_2(M)}) \approx 25$. For a 500~MHz bandwidth, this equates to $\tau_C \approx 50\mu s$). We compare the results for systems utilizing 4-QAM and 16-QAM signaling. 

There are several observations we would like to highlight from Fig.~\ref{fig:mod-order-comparison-product-codes}. First, in both the 4-QAM (solid lines) and 16-QAM (dashed lines) systems relying solely on designated training pilots to track the channel, the system performance hits a plateau. This leveling off is due to the fact that the channel decorrelates every $\sim 25$ codewords, but the training sequence is only sent every 100 codewords, which leads to stale channel estimates that fundamentally limit the achievable performance, regardless of the signal strength. In other words, the system ends up in States 2 and 3 from our analysis in Sec.~\ref{sec:analysis}. 

Secondly, for the 4-QAM case, all forms of additionally provided pilots substantially improve the BER performance to match that of the system with free pilots (i.e., the ideal scenario). This leads to more than 5~dB gain compared to the system relying solely on designated training pilots for a BER of $10^{-4}$. However, for the demodulator-provided pilot system there is a stark difference in its performance in the 4-QAM and the 16-QAM systems; for the 4-QAM scenario, the demodulator-provided pilots perform indistinguishably from the decoder-provided pilot and free pilot systems, while for the 16-QAM scenario, they perform decidedly worse than all other systems until the $E_b/N_0$ reaches 6~dB, when it drops below the system relying on the designated training sequences. 

This difference is due to the increased congestion of symbols in a 16-QAM constellation. As a result, it is much more likely that: (a) symbols are detected in error leading to an incorrect channel estimate, and (b) equalizing with incorrect channel estimates will cause symbols to be detected in error. In other words, we see that demodulator-provided pilots for higher modulation orders can enter a vicious cycle of error propagation that degrades the performance until the signal strength is high enough for HD detection to work well. This result points to the robustness that FEC provides decoder-provided pilots compared to DFEs; DFEs are more fragile at higher modulation orders, whereas the decoder-provided pilots are fortified by the decoding even as the modulation order increases. 

Comparing the performance of the CRC- and decoder-provided pilot systems in the 16-QAM case, in the region of extremely low $E_b/N_0$ (i.e., $E_b/N_0 < 2$), the CRC-provided pilots follow the performance of the designated training pilots while the decoder-provided pilots do a bit worse and follow the performance of the demodulator-provided pilots. As $E_b/N_0$ increases, both begin to perform better, but the CRC-provided pilots have a slight increase that quickly becomes negligible. Thus, depending on the larger implementation of the wireless system, either could be used with very similar performance. In the event that the CRC is not needed for higher layers of the protocol stack, it could be removed to increase the efficiency of the code. 

\begin{figure}[h!]
        \centering
        \includegraphics[width=0.9\linewidth]{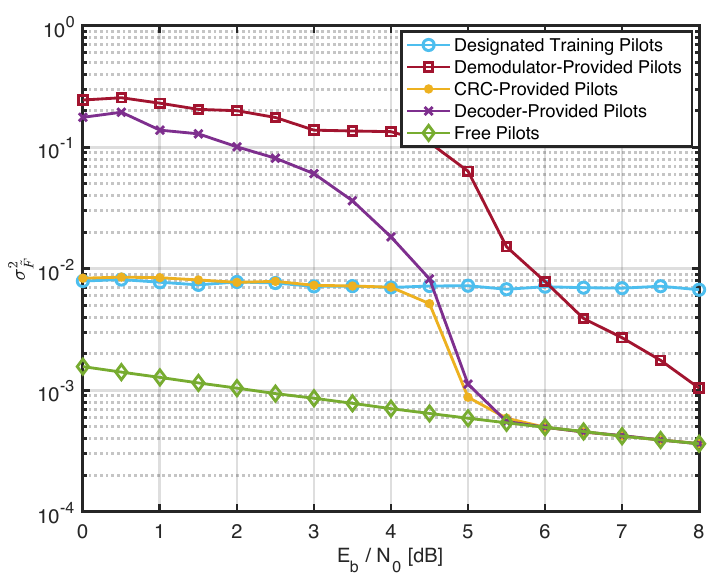}
        \caption{Variance of the channel estimation error for a system using $eBCH^2 [1024, 676]$ component codes with 11-bit CRC in a single-tap Rayleigh fading channel with $t_I = 100$ codewords and $D_I = 25$K symbols using 16-QAM signaling}
        \label{fig:chnl-variance-comparison-product-codes}
\end{figure}

\noindent\textit{Decoder-Provided Pilots Constantly Tracking the Channel}\\
To evaluate the ability of each proposed algorithm to track the time-varying channel, we calculate the channel estimation error for each data block for the 16-QAM system, and then plot the variance of the channel estimation error in Fig.~\ref{fig:chnl-variance-comparison-product-codes}. Following the trends observed in Fig.~\ref{fig:mod-order-comparison-product-codes}, the demodulator-provided pilots show the highest variance due to the high modulation order. In this plot, it is also easier to visualize the difference in performance between the CRC- and decoder-provided pilot systems; the decoder-provided pilot system performs worse than the CRC-provided system but better than the demodulator-provided system for low $E_b/N_0$ because the decoder corrects some errors which leads to a more accurate channel estimate than the demodulator provided system, but it has not corrected enough errors to avoid error propagation. 

The variance in channel estimation error of both the CRC- and decoder-provided systems decreases substantially to outperform the system relying only on designated training sequences when $E_b/N_0$ reaches about 4.5~dB. At this point the decoding is correct more often than not leading to improved channel estimation. 

\begin{figure}[h!]
        \centering
        \includegraphics[width=0.85\linewidth]{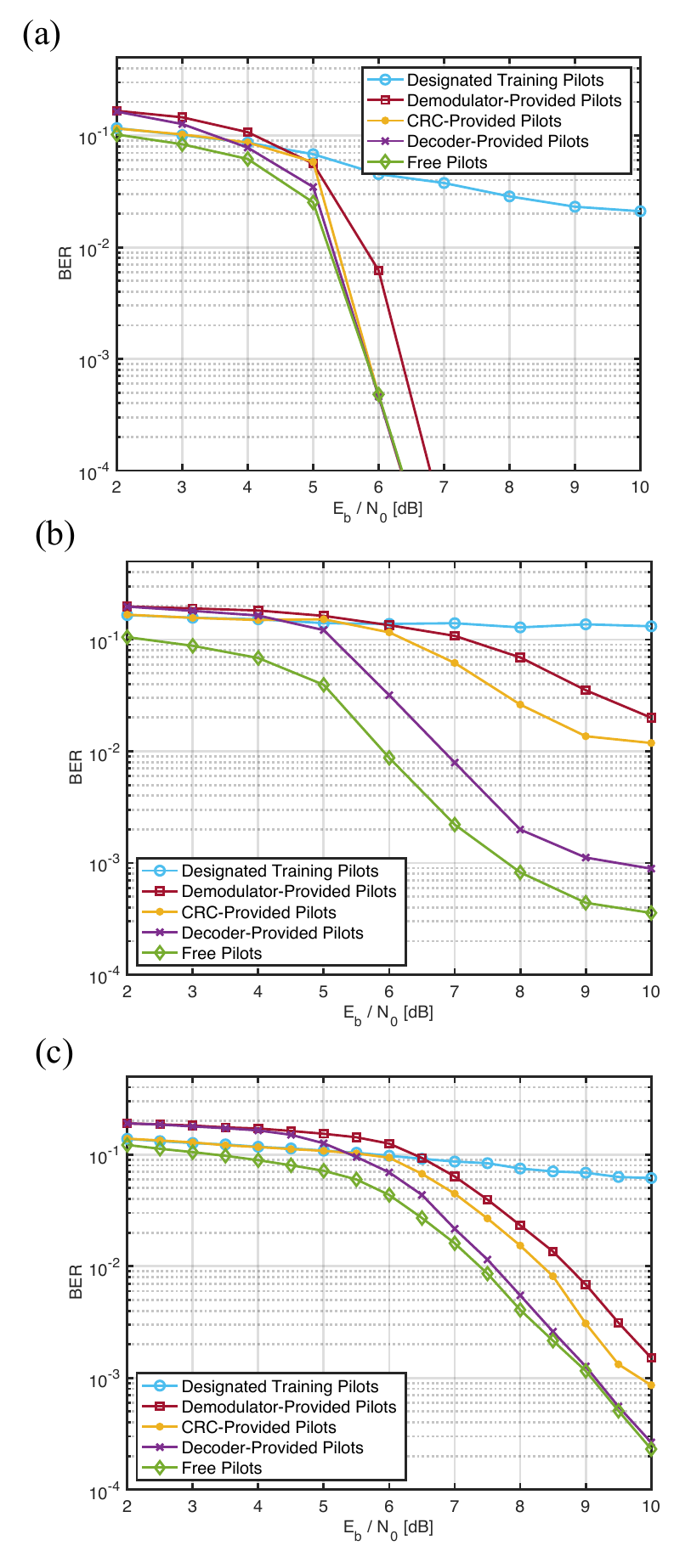}
        \caption{Bit error rate (BER) performance of a $eBCH^2 [1024, 676]$ component code with 11-bit CRC and 16-QAM signaling in a (a) Dicode channel with $\rho = 0.25$ $D_I = 10$K symbols; (b) Dicode channel with $\rho = 0.25$ $D_I = 1$K symbols; (c) Multi-tap channel with $\rho = [0.5, 0.25, 0.125]$ $D_I = 10$K symbols}
        \vspace{-0.5cm}
        \label{fig:chnl-comparison}
\end{figure}

\noindent\textit{Decoder-Provided Pilots Perform in Various Channels}\\
The channel characteristics will also impact the overall system performance and gains introduced by decoder-provided pilot systems. In Fig.~\ref{fig:chnl-comparison} we show the BER performance of a system using $eBCH^2 [1024, 676]$ component codes with 11-bit CRC and 16-QAM signaling in a variety of channels. Figure~\ref{fig:chnl-comparison}a shows the performance in the dicode channel described in Sec.~\ref{sec:results-framework} with $\rho = 0.25$ and the decorrelation interval $D_I = 10,000$ symbols, which corresponds to about 10 codewords. Figure~\ref{fig:chnl-comparison}b shows the performance in the same channel except the decorrelation interval $D_I = 1,000$ symbols, which corresponds to about 1 codeword, while Figure~\ref{fig:chnl-comparison}c shows the performance in the multi-tap channel with delay profile $D = [1, 0.5, 0.25, 0.125]$ with the decorrelation interval $D_I = 10,000$ symbols (i.e. about 10 codewords).
Comparing Fig.~\ref{fig:chnl-comparison}a and b, it is clear that the difference in the channels' coherence times leads to substantial differences in the achievable performance. In Fig.~\ref{fig:chnl-comparison}a, which has a channel that changes more slowly, the performance is similar to what was shown in Fig.~\ref{fig:mod-order-comparison-product-codes} for the single-tap channel. The performance of the decoder- and CRC-provided pilot systems quickly converge to that of the idea, free pilot system. However, in Fig.~\ref{fig:chnl-comparison}b, where the channel fluctuates much more quickly - decorrelating nearly every codeword - all the systems struggle. Even the free pilot system begins to plateau due to the speed of the channel changes with respect to the coherence time. 

In this case, the decoder-provided pilots thoroughly outperform the CRC-provided pilots because unlike in the slow fading channels used in Fig.~\ref{fig:chnl-comparison}a and Fig.~\ref{fig:mod-order-comparison-product-codes}, the channel is changing so fast that nearly perfect stale estimates used in the CRC-provided pilots system are less accurate than imperfect recent estimates from the decoder-provided pilots system. 

The decoder-provided pilots also outperform the CRC-provided pilots in the mid-range of $4 <E_b/N_0 < 5.5$ in Fig.~\ref{fig:chnl-variance-comparison-product-codes}a as well as in the multi-tap channel shown in Fig.~\ref{fig:chnl-comparison}c but by a smaller margin than in Fig.~\ref{fig:chnl-comparison}b. In general, for the multi-tap channel shown in Fig.~\ref{fig:chnl-comparison}c with a slower fading process, the performance is similar to that of Fig.~\ref{fig:chnl-comparison}a but with a more gradual slope of decline.

\begin{figure}[h!]
        \centering
        \includegraphics[width=0.9\linewidth]{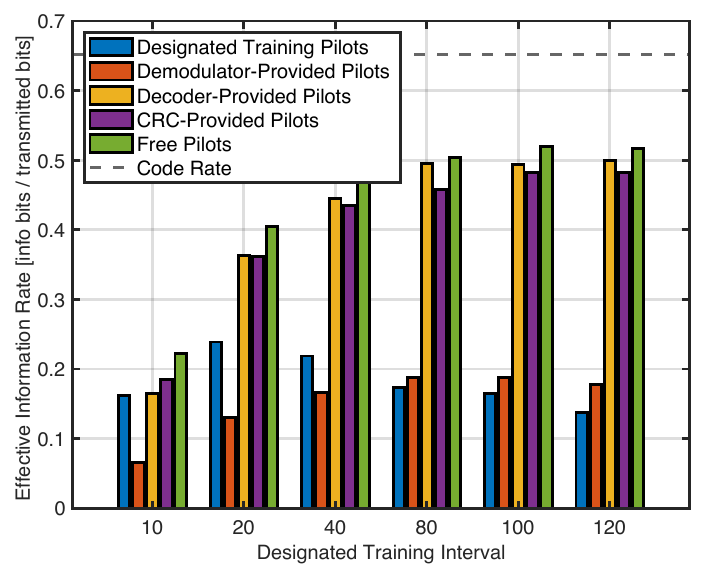}
        \caption{$eBCH^2 [1024, 676]$ with 11-bit CRC in a Dicode channel with $\rho = 0.25$ $D_I = 10$K symbols}
        \label{fig:eff-rate-product-codes}
\end{figure}

\noindent\textit{Decoder-Provided Pilots Increase Throughput}\\
Using Eq.~\eqref{eq:eff-rate}, we calculate the effective information rate for the same channel shown in Fig.~\ref{fig:chnl-comparison}a at $E_b/N_0= 5.5$~dB and plot the results in Fig.~\ref{fig:eff-rate-product-codes}. The x-axis represents different training intervals $t_I$. 

Considering the system relying only on the designated training pilots, shown in blue, the effective information rate peaks when $T_I \approx 20$ codewords and decreases with training intervals smaller or larger. This peak corresponds to when $t_I \approx \tau_{C}$. For the demodulator-provided pilots system, the performance is generally worse or comparable to that of the system without any additional pilots. Meanwhile, the decoder- and CRC- provided pilot systems strictly increase the throughput as the interval between designated training sequences decreases. This monotonic increase is due to the combination of the reduced training overhead as well as the improved BER performance. 

\begin{figure}[h!]
        \centering
        \includegraphics[width=0.9\linewidth]{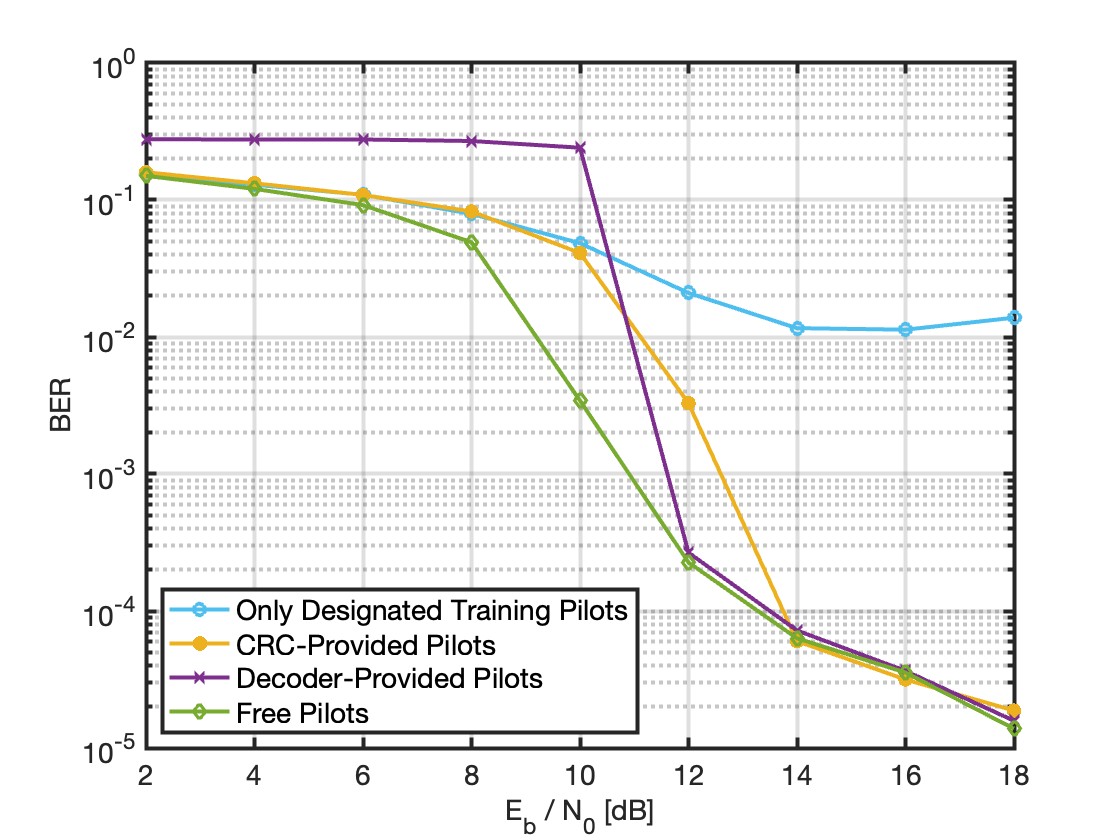}
        \caption{Bit error rate (BER) performance of a $eBCH^2 [1024, 676]$ code using 16-QAM OFDM with a short coherence time ($D_I = 25$ OFDM symbols) and a coherence bandwidth equal to the subcarrier spacing. We set the $t_I = 100$ OFDM symbols.}
        \vspace{-0.25cm}
        \label{fig:multi-carrier results}
\end{figure}

\noindent\textit{Decoder-Provided Pilots when Coding Across Frequency}\\
The prior analysis has considered single-carrier waveforms, but as mentioned in Sec.~\ref{sec:analysis-crc-provided}, we can easily extend our model to consider doubly-selective channels. Figure~\ref{fig:multi-carrier results} presents results of decoder- and CRC-provided pilots using 16-QAM OFDM signaling. In this case, we use an $eBCH^2 [1024, 676]$ code again with 256 subcarriers such that each OFDM symbol constitutes one codeword coded across frequencies. We assume a cyclic prefix eliminates any intersymbol interference and that each subcarrier experiences a Rayleigh channel independent of the other subcarriers' channels and has a decorrelation interval $D_I = 25$ symbols, and a training symbol is sent every 100 symbols. Due to the short coherence bandwidths, the channel is estimated individually for each subcarrier and extrapolated over time. 

The results in Fig.~\ref{fig:multi-carrier results} tell a similar story to those of the single-carrier systems. The BER of the system relying solely on the designated training symbols plateaus above $10^{-2}$, while the systems relying on decoder- and CRC-provided pilots are able to attain the best achievable performance. We see that the decoder-provided pilots can lead to worse error propagation in the low $E_b / N_0$ regime compared to the CRC-provided pilots. This difference in the error propagation is similar to what we observed for the single-carrier case, such as in Fig.~\ref{fig:mod-order-comparison-product-codes}. 

\begin{figure}[h!]
        \centering 
        \includegraphics[width=0.9\linewidth]{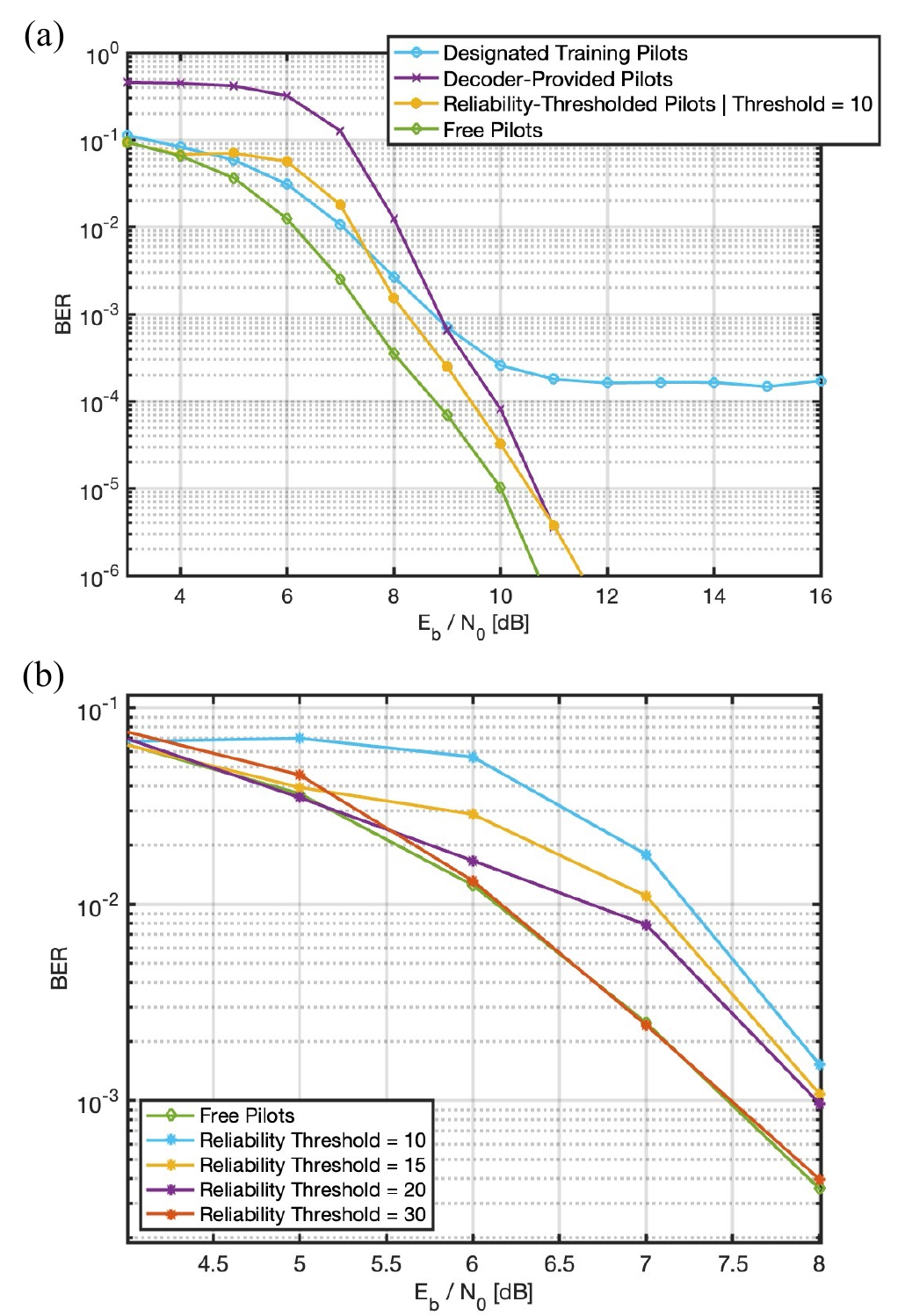}
        \caption{Bit error rate (BER) performance of a $eBCH [128, 113]$ code using BPSK OFDM with a short coherence time ($D_I = 10$ OFDM symbols) and a coherence bandwidth equal to the subcarrier spacing. We set the $t_I = 100$ OFDM symbols with decoder-provided pilots and reliability-thresholded pilots.} \vspace{-0.25cm}
        \label{fig:mc-so-results}
\end{figure}

\noindent\textit{Soft-Output Enhances Performance for Multi-Carrier Systems}\\
\label{sec:results-multicarrier-short-codes}
If soft-output is available for a multi-carrier waveform, we can use the reliability-thresholded decoder-provided pilots presented in Fig.~\ref{fig:system-model-long-codes}d. The BER performance of such a system is shown in Fig.~\ref{fig:mc-so-results}a, where $D_I = 10$ symbols. Designated training blocks are sent every $T_I = 100$ codewords, and we implement an eBCH [128, 113] code.  

The system relying solely on designated training symbols exhibits a plateau in performance above $10^{-4}$, similar to the plateaus observed in prior figures. The decoder-provided pilots allow for enhanced performance in the high $E_b / N_0$ regime. In the low $E_b / N_0$ regime, they lead to significant error propagation. 

In contrast, the reliability-thresholded decoder-provided pilots substantially reduce the amount of error propagation in the low $E_b / N_0$ regime, enabling performance closer to that of the ideal case. In Fig.~\ref{fig:mc-so-results}a, we use a reliability threshold of 10, and in Fig.~\ref{fig:mc-so-results}b, we show the BER results for the same system with different levels for the reliability threshold. As the threshold increases, the performance more tightly follows that of the ideal scenario. 

\begin{figure}[h!]
        \centering \includegraphics[width=0.9\linewidth]{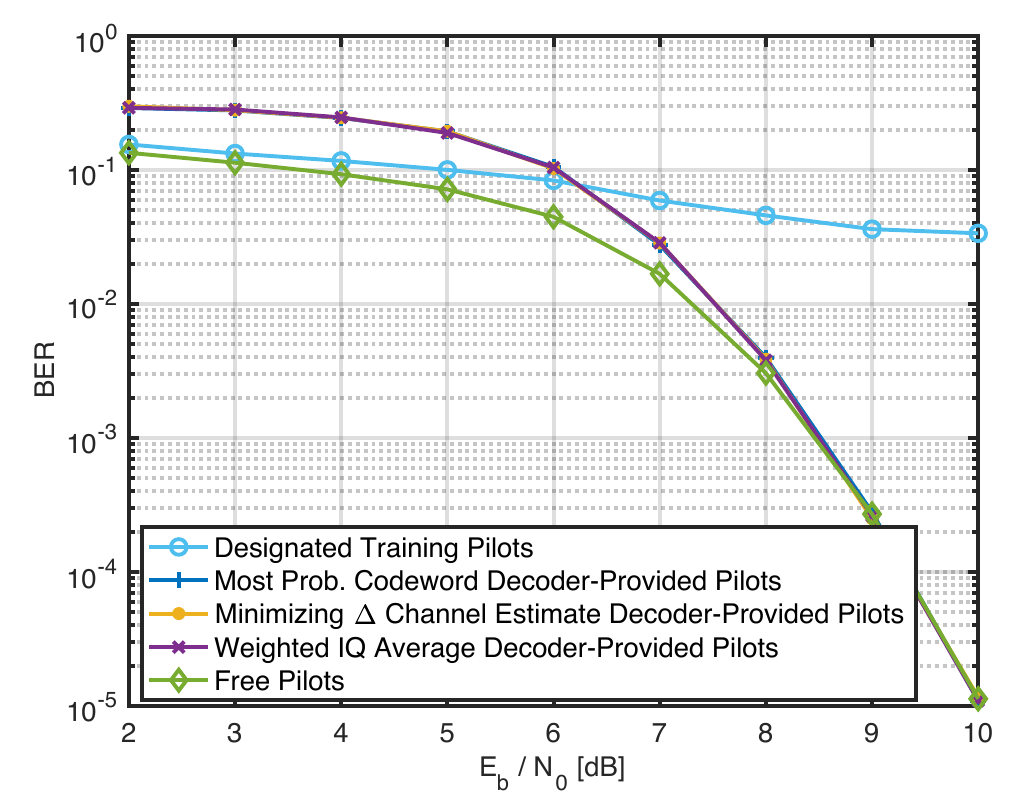}
        \caption{Bit error rate (BER) performance of an $eBCH [128, 113]$ code in a single-tap Rayleigh fading channel with $t_I = 100$ codewords and $D_I = 1$K symbols using 16-QAM signaling.}
        \vspace{-0.25cm}
        \label{fig:initial-results-short-codes}
\end{figure}

\noindent\textit{Hard-Decision Enables Best Achievable Performance when Coding Across Time}\\
\label{sec:results-short-codes}
In Fig.~\ref{fig:initial-results-short-codes} we show the results utilizing the soft-output decoding variations of decoder-provided pilots described in Sec.~\ref{sec:system-model} assuming a list decoder. We implement a single-tap Rayleigh fading channel with $t_I = 100$ codewords and $D_I = 1$K symbols, which corresponds to about 8 codewords for the $eBCH [128, 113]$ code utilized. 

In the region of low signal strength (i.e., $E_b /N_0 < 6$), the decoder-provided pilot systems do worse than the system relying solely on designated training pilots. In other words, in this region the systems all experience error propagation due to incorrect channel estimates. As the $E_b/N_0$ continues to increase, however, the decoder-provided systems surpass the performance of the baseline system to approach the optimal performance. 

There is also no visible difference in performance between the systems utilizing the soft-output to enhance channel estimation (i.e., the ``Weighted IQ Average Provided Pilots" shown in purple and the ``Minimizing $\Delta$ Channel Provided Pilots" shown in yellow) and the system not taking advantage of soft-output information (i.e., the ``Most Probable Codeword Provided Pilots" shown in blue). Thus, when coding across time, optimal performance is reached using hard-decision outputs alone. 

\noindent\textit{Performance of Decoder-Provided Pilots with Non-Systematic Codes}\\
\label{sec:results-sys-v-nonsys}
Beyond the number of errors found in the decoding, how these errors propagate will also depend on the structure of the code used. In general, error correcting codes fall into two categories: systematic and non-systematic. Systematic codes (like the eBCH codes used in our prior analysis) are structured such that the information bits can be pulled directly from the codeword. In other words, a codeword is generated simply by adding redundant bits to the existing information bits according to some generator function or matrix.  

For non-systematic codes, the information bits are not in the codeword; they are found by solving a set of equations with the codeword as input. This distinction means that for a systematic code, if there are $e$ bit errors observed within an $n$-length codeword, there will be $e$ or fewer bit errors within the $k$-length information sequence. However, for a non-systematic code, each $e$ bit error will contribute to errors in solving the system of equations to find the $k$-length information sequence; therefore, the number of information bits received in error will only be bounded by $k$. In fact in many scenarios $e$ bit errors in the detected codeword will often lead to $\geq e$ errors in the $k$-length information sequence.

\begin{figure}[h!]
        \centering
        \includegraphics[width=0.9\linewidth]{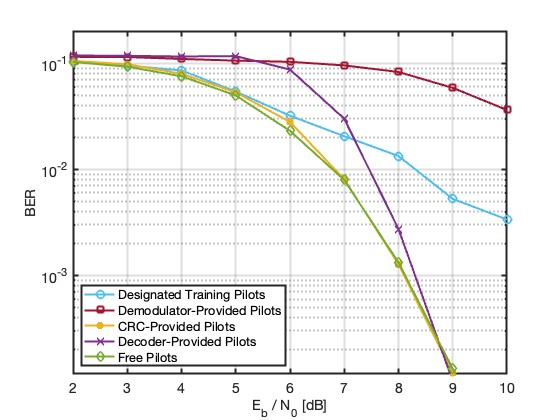}
        \caption{Bit error rate (BER) performance of the $[128, 54]$ CA Polar code in 5G NR in a single-tap Rayleigh channel with $t_I = 100$ codewords, $D_I = 1$K symbols, and 16-QAM signaling.}\vspace{-0.25cm}
        \label{fig:ca-polar-code results}
\end{figure}

In this analysis we present decoder-provided and CRC-provided pilots using a non-systematic code, namely the CRC-assisted (CA) Polar code used in 5G-NR with 30 information bits appended with a 24-bit CRC before the non-systematic encoding that maps the 54-bit sequence to a 128-bit codeword $[128, 54]$ according to~\cite{bioglio2020design}. The CA successive cancellation list (SCL) decoder is used with a list size of 8, and the channel is single-tap and Rayleigh fading with $t_I = 100$ codewords and $D_I = 1$K symbols. 16-QAM signaling is used. The BER performance is shown in Fig.~\ref{fig:ca-polar-code results}.

Similar to the results from Fig.~\ref{fig:mod-order-comparison-product-codes}, the system using demodulator-provided pilots suffers from severe error propagation, which leads to BERs higher than that of the system relying only on designated training pilots, even as the $E_b / N_0$ increases. The decoder-provided pilot system shown in purple also experiences severe error propagation that leads to performance on par with the demodulator-provided pilot system until $E_b / N_0$ reaches about 7~dB, where it drops to approach the optimal performance of the system provided with free pilots. Meanwhile, the CRC-gated decoder-provided pilots, shown in yellow, allow for optimal and close-to-optimal performance throughout the displayed range of $E_b / N_0$ values. 

Compared with the systematic $eBCH^2$ code from Fig.~\ref{fig:mod-order-comparison-product-codes}, this difference in performance is more pronounced due added error propagationn of the non-systematic code\footnote{One thing to note is that we assume our algorithm is code agnostic. If the decoder-provided pilot algorithm were to be designed with knowledge that the code is non-systematic, the non-systematic decoder need not propagate errors.}.

\subsection{Comparing Long and Short Code Performance Channels with Different Coherence Times}
\label{sec:results-long-v-short}
\begin{figure}[h!]
        \centering
        \includegraphics[width=0.9\linewidth]{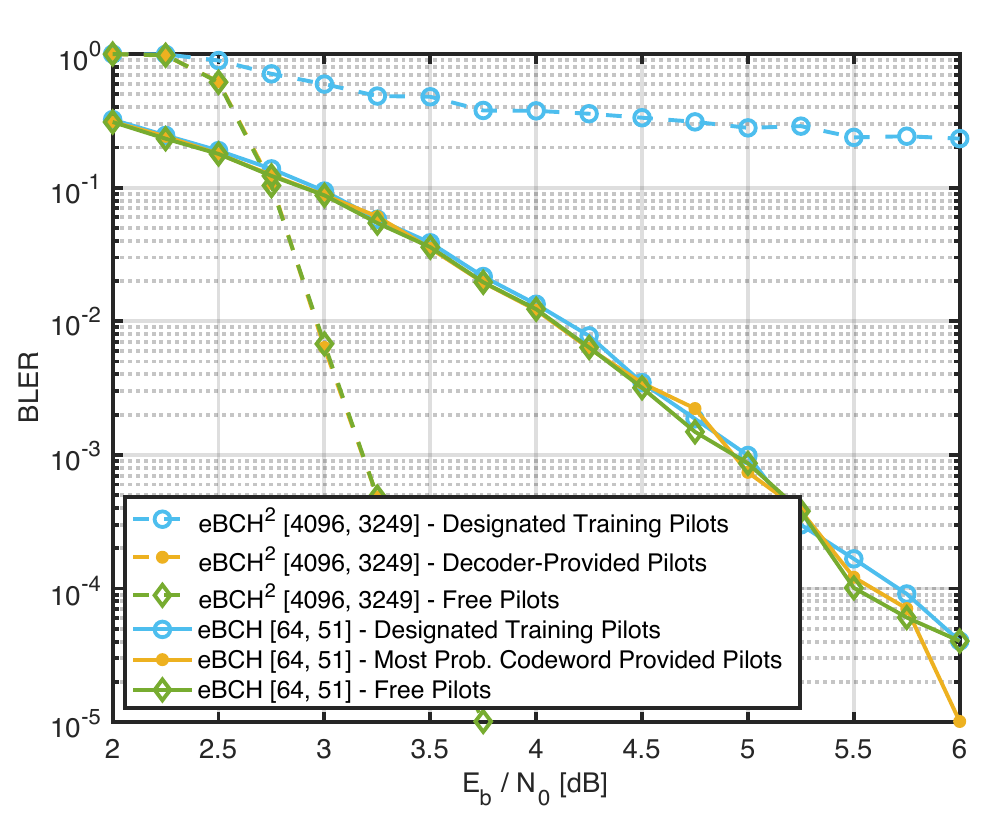}
        \caption{Block Error Rate (BLER) performance using BPSK signaling with $t_I = 100$ codewords and $D_I = 10k$ symbols}\vspace{-0.25cm}
        \label{fig:long-v-short-10k}
\end{figure}

\begin{figure}[h!]
        \centering
        \includegraphics[width=0.9\linewidth]{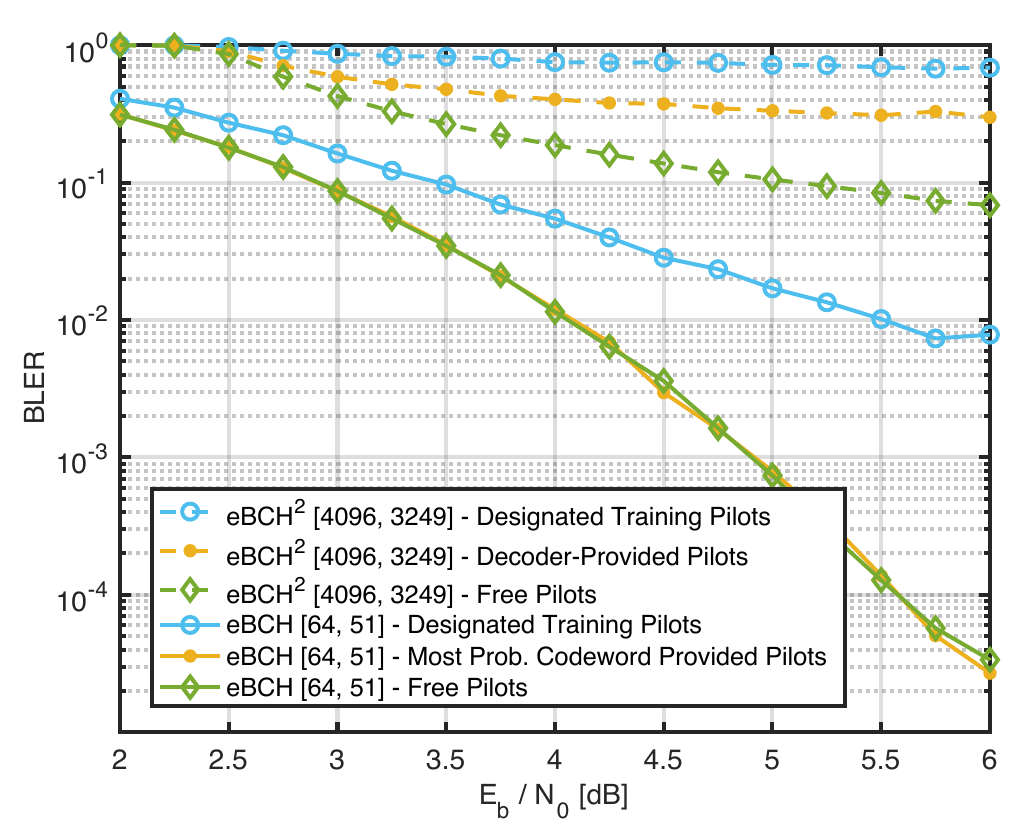}
        \caption{BLER performance using BPSK signaling with $t_I = 100$ codewords and $D_I = 1k$ symbols}\vspace{-0.25cm}
        \label{fig:long-v-short-1k}
\end{figure}

Although the soft-output information does not prove particularly useful for decoder-provided pilots when coding across time, as indicated by the analysis presented in Sec.~\ref{sec:analysis-limit-length}, there are still instances when the channel might motivate using shorter versus longer codes. We demonstrate this through simulations in Fig.~\ref{fig:long-v-short-10k} and~\ref{fig:long-v-short-1k}, where we plot the block error rate (BLER) performance of two different codes in two different channels. Results using $eBCH^2 [64, 57]^2 = [4096,3249]$ component codes with BPSK signaling are shown by dashed lines. The training interval $t_I = 100$ codewords. Meanwhile, results using the list decoder on $eBCH [64,51]$ codes with BPSK signaling and the same training interval of $t_I = 100$ codewords are shown by solid lines.

In Fig.~\ref{fig:long-v-short-10k}, we demonstrate the performance in a single-tap Rayleigh channel with a decorrelation interval of $D_I = 10,000$ symbols. For the product code, this corresponds to about 9 codewords, while for the shorter code it corresponds to about 156 codewords. In Fig.~\ref{fig:long-v-short-1k}, we use a single-tap Rayleigh channel with a shorter decorrelation interval of $D_I = 1,000$ symbols, which is less than 1 codeword for the product code and is about 15 codewords for the shorter code. 

 Note that these codes have almost the same rates: 0.793 for the product code and 0.797 for the shorter code. Thus according to~\cite{shannon1948mathematical}, we would anticipate the longer code to perform better, and in the slow fading channel we see just that; in Fig.~\ref{fig:long-v-short-10k}, the decoder-provided pilots are able to improve the performance substantially leading to a steep drop in the BLER. In contrast, the performance of the shorter code in the same channel shown by the solid lines falls off much more gradually. It is also interesting to note that the system relying solely on designated training pilots is able to reach the optimal performance because the channel is changing slowly enough for the training interval to keep up. 

However, in the fast fading channel, the product code does exceptionally poorly, as illustrated in Fig.~\ref{fig:long-v-short-1k}. Even the best achievable performance indicated by the ``Free Pilots" shown in green begins to level off well above $10^{-2}$. This is because the length of the codeword is so long that the channel decorrelates multiple times within a codeword, making accurate channel estimation for the given system impossible. For the shorter code in the fast fading channel in solid lines, the system relying solely on designated training pilots begins to struggle because the training interval is now less than the decorrelation interval of the channel. However, the decoder- and demodulated provided pilots are still able to substantially improve the performance of the system. This follows the expectation provided by our analysis in Sec.~\ref{sec:analysis-limit-length}; in fast fading channels shorter codes can be ideal to ensure the system approaches the highest achievable capacity. 

\section{Discussions \& Conclusion}
\label{sec:conclusion}
In this paper, we present the use of decoded data blocks as additional pilot training sequences in time-varying scenarios. Unlike previous approaches to joint channel estimation and error correction, our approach is non-iterative for low-latency applications and code-/decoder-agnostic for flexible deployment. It does not require soft-output information, but can provide gains by using hard-decisions based on soft-output information. We demonstrate the performance analytically and through simulations show that decoder-provided pilots can increase the throughput of systems by simultaneously increasing the data rate and reducing the required transmission overhead for channel training. We also show that short codes can be preferable in fast fading channels to avoid stale channel estimates in the single-carrier case. Soft-output information is also demonstrated to improve the performance when coding across frequency. 

We would also like to emphasize that this approach can enhance any waveform to allow for the simultaneous channel tracking and data transmission. Furthermore, unlike data-aided and DFE approaches the decoder-provided pilots give a single-shot measurement for low-latency and dynamic settings, while maintaining robustness even when high modulation orders are used. 

For our analysis, we assume an auto-regressive channel model because many time-varying channels can be represented as such, but we would expect this approach to be robust when applied to other channels as well. In essence, decoder-provided pilots allow for near-continuous channel probing, reducing the system's reliance on accurate channel modeling. If a channel model is known and pilot symbols are used, approaches like Kalman filtering could be implemented. However, unlike Kalman filtering, decoder-provided pilots do not require prior channel knowledge of the channel model and can eliminate the need for designated pilot sequences altogether (given proper choice of the code rate and length). 

Future studies for decoder-provided pilots could explore the improvements of implementing Kalman filtering with decoder-provided pilots. 

\bibliographystyle{IEEEtran}
\bibliography{bibliography.bib}

\end{document}